\documentclass[12pt]{amsart}
\usepackage{amsmath,amssymb,amsfonts}
\usepackage{epsfig}

\newcommand{\res[1]}{\operatornamewithlimits{Res}\limits_{#1}}

\newtheorem{theorem}{Theorem}[section]
\newtheorem{proposition}[theorem]{Proposition}
\newtheorem{lemma}[theorem]{Lemma}
\newtheorem{corollary}[theorem]{Corollary}
\newtheorem{remark}[theorem]{Remark}
\newfont{\tenbi}{cmbxti10}

\def\la {\lambda}

\renewcommand{\Re}{\mathop{\mathrm{Re}}}

\begin{document}
\title [Periodic billiard orbits on $n$--dimensional
ellipsoids ...]{Periodic billiard orbits on $n$--dimensional
ellipsoids with impacts on confocal quadrics and isoperiodic deformations.}

\author{Simonetta Abenda}
\address{Dipartimento di Matematica e CIRAM,
Universit\`a degli Studi di Bologna, Italy,
abenda@ciram.unibo.it}
\author{Petr G. Grinevich}
\address{L.D.Landau Institute for Theoretical Physics,
pr. Ak Semenova 1a, Chernogolovka, 142432, Russia,
{\footnotesize pgg@landau.ac.ru}}
\thanks{
This work has been partially supported by PRIN2006 ''Metodi
geometrici nella teoria delle onde non lineari ed applicazioni'', by the
Program of Russian Academy of Sciences ``Fundamental problems of 
nonlinear dynamics'', RFBR grant 09-01-92439-KE-a and by the grant 
NSh.1824.2008.1 -- Leading Scientific Schools of the Russian 
Federation Presedent's Council for grants}

\begin{abstract}
In our paper we study periodic geodesic motion on multidimensional 
ellipsoids with elastic impacts along confocal quadrics. We show that 
the method of isoperiodic deformation is applicable.
\end{abstract}
\maketitle

\section{Introduction}

One of the classical integrable problems is the geodesic motion on
the ellipsoid. Complete integrability  of the 3-dimensional
problem was proven by Jacobi \cite{Jacobi}. Explicit formulas in
terms of genus 2 $\theta$-functions were found by Weierstrass
\cite{Weier}. Integrability of the multidimensional problem was
established by Moser \cite{Moser}. Connections with the Neumann
problem were exploited by Kn\"orrer \cite{Knorr2}. Integrability
of the Neumann problem has a very transparent explanation, based
on the following observation by Trubowitz and Moser (see
\cite{Moser}): taking properly normalized eigenfunctions at the
right ends of gaps for the $(N-1)$-gap 1-dimensional stationary
Schr\"odinger operator we obtain the coordinates of the particles.

The isomorphism between the geodesics on the ellipsoid and the
Neumann problem is nontrivial, because it includes a nonlinear
change of independent variable. The motion on the Jacobian for the
Neumann problem is uniform, and the motion for the geodesics on
the ellipsoid is along straight lines, but the speed in the
natural parameter is not constant. It means, that the analog of
the Trubowitz-Moser isomorphism maps this dynamics into the
Harry-Dym equation instead of the Korteweg-de Vries (KdV)
equation. On the level of the associated spectral problems it
means, that one has to consider the 1-dimensional acoustic problem
\begin{equation}
\label{eq:acoustic}
-\partial^2_t \Psi(E,t) = E u(t)  \Psi(E,t), \quad E=1/\lambda
\end{equation}
instead of the 1-dimensional Schr\"odinger problem
\begin{equation}
\label{eq:schrodinger}
-\partial^2_\sigma \Psi(E,\sigma) + \tilde u(\sigma)  \Psi(E,\sigma) = E \Psi(E,\sigma).
\end{equation}
The connection between the the geodesics on the ellipsoid and the
Harry-Dym equation was first observed by Cao \cite{Cao}, and
independently by Veselov \cite{Ves3}, \cite{Ves4}.

From the geodesic motion on the ellipsoid one can construct an
integrable billiard motion taking the
geodesic dynamics on the $n$-dimensional ellipsoid
\begin{equation}
\label{eq:ellipsoid}
Q = \left\{ \frac{x_1^2}{a_1}+\cdots +\frac{x_{n+1}^2}{a_{n+1}}=1
\right\},
\end{equation}
combined with elastic reflections along one or more confocal quadrics
$$
Q_{d_s}= \left\{ \frac{x_1^2}{a_1-d_s}+\cdots
+\frac{x_{n+1}^2}{a_{n+1}-d_s}=1 \right\}, \ \ s=1,\ldots,r.
$$
$d_s$ being arbitrary fixed parameters. An amazing property of
this motion is that all conservation laws remain invariant after
reflections, therefore this dynamical system is also completely
integrable. Such property was first observed by Chang and Shi
\cite{Chang} for the triaxial ellipsoids, while its
multidimensional generalization was obtained in  \cite{Chang0}. As
a consequence the trajectories between impacts (arches of
geodesic) or their continuations are tangent to the same set of
$n-1$ quadrics (caustics) in ${\mathbb R}^{n+1}$ that are confocal
to $ Q$ and $Q_{d_S}$. In the same papers it was proven the
Poncelet property: if one trajectory is closed after $p$ bounces,
all the trajectories sharing the same constants of motion are also
closed after $p$ bounces and have the same length.

The billiard on the ellipsoid is described by the maps
$\mathcal{B}_s\, : (x^{(N)} ,v^{(N)}_{\rm{out}})\mapsto
(x^{(N+1)}, v^{(N+1)}_{\rm{out}})$, where $x^{(N)},v^{(N)}_{\rm{out}}\in
{\mathbb R}^{n+1}$, $N\in \mathbb N$, are respectively the
coordinates of the $N$-th impact point on $ Q\cap Q_{d_s}$ and the
outgoing velocity at this point, whereas $x^{(N+1)},v^{(N+1)}_{\rm{out}}$
denote the same objects at the next impact point. This billiard map is
transcendental in the following sense: the shift takes place on a
non-Abelian variety and, as a result, it is not fixed. This property was first
observed by Veselov \cite{Ves2} for the billiard on triaxial
ellipsoids and generalized for higher dimensions by Abenda and Fedorov
in \cite{AF2}. More precisely, the generic complex invariant manifolds
are not Abelian varieties, but open subsets of theta-divisors
of $n$-dimensional hyperelliptic Jacobians of a given hyperelliptic
curve $\Gamma$. Thus, in the real domain, the new
values $(x^{(N+1)},v^{(N+1)}_{\rm{out}})$ are determined
by solving a system of transcendental equations which involve the
inversion of hyperelliptic integrals, whereas in the complex
coordinates the billiard map is infinitely many valued.

We stress that all of these properties stand in sharp contrast to
both the Birkhoff ($p=0$) \cite{Birk,MosVes,Ves,Ves1} and the
Hooke ($p=2$) \cite{Fed2} billiard systems, where the billiard map
is given by a shift by a constant vector on a complex invariant
manifold which is an open subset of an Abelian variety. Due to the above
properties, the billiard maps $\mathcal{B}_s$ can be regarded as a
discrete analog of the algebraic integrable systems with
deficiency studied, in particular, in \cite{Van,AF,AlberFed,ERP}).
The latter systems, although being Liouville integrable, are not
algebraic completely integrable: their generic complex solutions
have movable algebraic branch points and are single-valued on an
infinitely sheeted ramified covering of the complex plane $t$, so
they possess the so called weak Painlev\'e property. The branching
of the billiard maps $\mathcal{B}_s$ can be viewed as a discrete version
of such a property (see also \cite{Gramm}).

The development of the finite-gap approach to the 1-d Schr\"odinger
operator (\ref{eq:schrodinger}) and the KdV equation was started by Novikov
\cite{NOV}. Generic finite-gap potentials are quasiperiodic. Selection
of purely periodic potentials is rather nontrivial.

The principal periodicity constraints discussed in the literature
are the following:
\begin{enumerate}
\item Periodicity in one real variable.
\item Periodicity in two real variables.
\item Double periodicity in one complex variable. Such solutions are
  called elliptic solitons.
\end{enumerate}

The problem of selecting space-periodic potentials for
the 1-D Schr\"odinger operator was solved by Marchenko
and Ostrovski \cite{Marchenko} in terms of conformal maps. It is also
possible to study the variety of all spectral curves generating purely
periodic potentials, using the so-called isoperiodic deformations (they
can be interpreted as Loewner equations for these conformal maps).
Period preserving deformations first arose in the papers by Ercolani,
Forest, McLaughlin and  Sinha \cite{Ercolani}, Krichever \cite{Krichever3},
Schmidt \cite{Schmidt}. Grinevich and Schmidt \cite{Grinevich} suggested to
use isoperiodic deformations for selecting pure periodic solutions,
they also showed that these flows can be transformed to a very simple
algebraic form by extending the phase space.

The finite-gap theory for the acoustic problem (\ref{eq:acoustic})
was developed by Dmitrieva and collaborators \cite{Dmitrieva1},
\cite{Dmitrieva2}, \cite{Dmitrieva3}. This problem can be
transformed to the Schr\"odinger problem (\ref{eq:schrodinger}) by
a reciprocal transformation over the independent variable.
Equations for the isoperiodic deformations of this problem were
written by D.Zakharov \cite{Zakh}. These results can be
automatically applied to the periodic geodesics.

The first non-trivial examples of elliptic solitons for the KdV
equation (and 1-D Schr\"odinger operators potentials respectively)
were found by Dubrovin and Novikov \cite{Dub_Nov2}.
These solutions correspond to genus $g=2$ curves. Airault, McKean and
Moser \cite{AMM77} established the connection of KdV elliptic solitons
with a special reduction of the elliptic Calogero-Moser system valid
for arbitrary genera. An analogous connection between the
Kadomtsev-Petviashvili (KP) equation and the full elliptic Calogero-Moser
system was discovered by Krichever \cite{Krich}. A complete
characterization of the double-periodic in the complex $x$-variable KdV
and KP solutions in terms of so-called tangential coverings was
obtained by Treibich and Verdier \cite{TV}, \cite{Tr}. Due to the
Trubowitz-Moser isomorphism these results can be also applied to
fully characterize and construct real geodesics and real billiard
trajectories, which are double-periodic in the complex length
\cite{Fed05}, \cite{AF2}, \cite{A}.

It is natural to pose the problem of constructing real billiard
trajectories on $n$-dimensional ellipsoids which are
simply-periodic. In our paper we show, that the isoperiodic
technique can be applied to this problem.

\section{Geodesics on Ellipsoids}

In this section, to settle notations, we briefly outline the
description of the geodesics and billiard motions on a quadric.
The geodesic motion on an $n$-dimensional ellipsoid $Q$ is well
known to be integrable and to be linearized on a covering of the
Jacobian of a genus $n$ hyperelliptic curve \cite{Moser}. Namely,
let $t$ be the natural parameter of the geodesic and
$\la_1,\dots,\la_n$ be the ellipsoidal coordinates on $ Q$ defined
by the formulas
\begin{equation}
\label{spheroconic22} x_i=\sqrt{ \frac{a_i(a_i-{\la}_1)\cdots
(a_i-{\la}_n)} { \prod\limits_{j\ne i}(a_i-a_j)} }, \qquad i=1,\dots,
n+1.
\end{equation}
Then, denoting $\dot\lambda_k= d\lambda_k/d t$ the corresponding
velocities, the total energy $\displaystyle \frac 12 (\dot x, \dot
x)$ takes the St\"ackel form
\begin{gather*}
H = -\frac{1}{8}\sum^{n}_{k=1} \left[ \frac
{\prod\limits^{n}_{j\ne k} (\lambda_{k}-\lambda_{j})}
{\prod\limits^{n+1}_{j=1}(\lambda_k-a_{j})} \right] \lambda_k
\dot\lambda^{2}_{k}.
\end{gather*}
According to the St\"ackel theorem, the system is Liouville
integrable. After fixing the constant of motion $c_k$,
$k=1,\ldots,n-1$ and assuming $H=1/2$ we obtain Dubrovin equations:
\begin{equation}
\label{eq:dubrovin} \dot \la_k =
-2\frac{w_k}{\la_k\prod\limits^{n}_{j\ne k}
(\lambda_{k}-\lambda_{j})}.
\end{equation}
Here
\begin{equation}
w_k^2 =-\la_k  (\la_k-a_1)\cdots (\la_k-a_{n+1})
(\la_k-c_1)\cdots (\la_k-c_{n-1}), \ \ \ \ k=1,\ldots,n.
\end{equation}
The evolution of the divisor $\la_k$ is defined on the  genus $n$
hyperelliptic curve
\begin{equation}\label{gamma}
\Gamma \; : \;\; \big\{w^2= -\lambda \prod_{i=1}^{n+1} (\lambda
- a_i) \prod_{k=1}^{n-1} (\lambda-c_k) \big\} = \big\{w^2=
-\prod_{i=0}^{2n} (\lambda - b_i) \big\} ,
\end{equation}
where we set the following notation throughout the paper
\begin{equation} \label{rampts}
\{ \, 0, a_1<\dots< a_{n+1}, c_1<\cdots< c_{n-1} \, \} = \{ b_0=0<
b_1< \cdots< b_{2n} \}.
\end{equation}

The constants of motion $c_k$ have the following geometrical
meaning (see \cite{Moser}): the corresponding geodesics are
tangent to the quadrics $Q_{c_1},\dots , Q_{c_{n-1}}$ of the
confocal family $Q_c = \left\{ x_1^2/(a_1-c) +\dots +
x_{n+1}^2/(a_{n+1}-c) =1 \right\}$.

In particular, the reality condition for geodesics on ellipsoids
is equivalent to either $c_i = b_{2i}$ or $c_i = b_{2i+1}$,
$i=1,\dots,n-1$ (see \cite{Knorr,Audin}). Since we are interested
in the reality problem, we take,without loss of generality,
$b_{2i-1}<\lambda_i<b_{2i}$, $i=1,\dots, n$.

Let us recall (see \cite{Cao}, \cite{Ves3}) how to reconstruct the
geodesic motion from the solutions of the acoustic problem
(\ref{eq:acoustic}). The $n$-gap eigenfunctions of (\ref{eq:acoustic})
are defined by the following properties:
\begin{enumerate}
\item For any $t$, $\Psi(\gamma,t)$ is meromorphic in $\gamma$ on
the
  spectral curve $\Gamma\backslash\{\lambda=0\}$.
\item For any $t$, $\Psi(\gamma,t)$ has $n$ simple poles at the points
  $\gamma_1(0)$,\ldots,$\gamma_n(0)$. Here $\la_j(t)$ is the
  projection of  $\gamma_j(t)$ to the $\la$-plane.
\item The there exists a function $\sigma(t)$ such, that
$$
\exp{\left(-\frac{i\sigma(t)}{\sqrt{\la}}\right)}\Psi(\gamma,t)
$$
is regular near the point $\la=0$.
\item
\begin{equation}
\label{eq:psias}
\Psi(\lambda,t)=1 +\frac{it}{\sqrt{\la}}+ O\left(\frac{1}{\la}\right)
\quad \mbox{as} \quad \la\rightarrow\infty.
\end{equation}
\end{enumerate}
Let us define
\begin{equation}
x_j(t)=\Psi(a_j,t)
\sqrt{\frac{a_j\prod\limits_{k=1}^{n}(a_j-\la_k(0))}
{\prod\limits_{k\ne j}(a_j-a_k)}}
\end{equation}
Consider the following pair of meromorphic forms in the
$\la$-plane:
\begin{gather}
\Omega_1=\Psi(\gamma,t)\Psi(\tau\gamma,t)
\frac{\prod\limits_{k=1}^{n}(\la-\la_k(0))}
{\prod\limits_{k=1}^{n+1}(\la-a_k)}d\la,\\
\Omega_2=\dot\Psi(\gamma,t)\dot\Psi(\tau\gamma,t)
\frac{\la\prod\limits_{k=1}^{n}(\la-\la_k(0))}
{\prod\limits_{k=1}^{n+1}(\la-a_k)}d\la.
\end{gather}
Here $\tau$ denotes the hyperelliptic involution. From
(\ref{eq:psias}) it follows, that the integrals of $\Omega_1$,
$\Omega_2$ over sufficiently large contour are equal to $2\pi i$.
The residues of $\Omega_1$ and $\Omega_2$ at the point $a_j$ are
$\frac{x_j^2(t)}{a_j}$ and $\dot x_j^2(t)$ respectively. Therefore
we obtain (\ref{eq:ellipsoid}) together with the normalization
condition:
\begin{equation}
\sum\limits_{j=1}^{n+1} \dot x_j^2(t)=1.
\end{equation}
The motion is governed by the Lagrange multipliers principle:
$$
\ddot x_j(t) = -u(t) \frac{x_j(t)}{a_j}
$$

After the re-parameterization
\begin{equation} \label{tau-1}
dt =\frac{\la_1\cdots\la_n}{\sqrt{b_1b_2\ldots b_{2n}}}\, d\sigma ,
\end{equation}
(this re-parameterization coincides with the change of independent
variables, connecting the acoustic operator (\ref{eq:acoustic})
with Schr\"odinger operator (\ref{eq:schrodinger})) the evolution
of $\lambda_{i}$ is described by quadratures
\begin{gather}
\sqrt{\prod\limits_{j=1}^{2g} b_j} \cdot \left[\frac{\la_1^{k-1}}{2 w_1} d\la_1
+\cdots + \frac{\la_n^{k-1} }{2 w_n}d\la_n \right]
 =\Bigg \{
\begin{aligned}
d\sigma\quad \mbox{ for } & k=1\, , \\
0\quad \mbox{ for } & k=2, \dots, n ,
\end{aligned}
\label{quad2}
\end{gather}

The quadratures involve $n$ independent holomorphic differentials
on $\Gamma$.

\noindent
Let $\alpha_i,\beta_i$, $i=1,\dots, n$ be the
conventional homological basis depicted in Figure 1 and
\begin{equation}\label{basis}
\omega_j =\sqrt{\prod\limits_{j=1}^{2g} b_j} \cdot 
\frac{\la^{j-1} d\la }{w}, \quad\quad j=1,\dots, n,
\end{equation}
be the usual basis of holomorphic differentials. (\ref{quad2})
give rise to the Abel--Jacobi map of the $n$-th symmetric product
$\Gamma^{(n)}$ to the Jacobian variety of $\Gamma$ (see for example
\cite{Gunn}):
\begin{equation}\label{AB}
u_j=\int \limits^{P_1}_{P_0} \omega_j + \cdots + \int
\limits^{P_n}_{P_0} \omega_j =\Bigg \{
\begin{aligned}
2\sigma + const.,\quad \mbox{ for } & j=1\, , \\
const., \quad \mbox{ for } & j=2, \dots, n ,
\end{aligned} \end{equation}
where and $P_0$ is a fixed basepoint and $ P_k=\left(\lambda_k,w_k
\right) \in \Gamma$, $k=1,\dots, n$.  Then, the geodesic motion in
the new parameterization is linearized on the Jacobian variety of
$\Gamma$. Its complete theta-functional solution was presented in
\cite{Weier} ($n=2$), and in \cite{Knorr} ($n>2$), whereas a
topological classification of real geodesics on quadrics was made
in \cite{Audin} and an algebraic geometric characterization of a
dense set of real closed geodesics on ellipsoids has been given in
\cite{A}.

\begin{center}
{\epsfxsize=14cm \epsffile{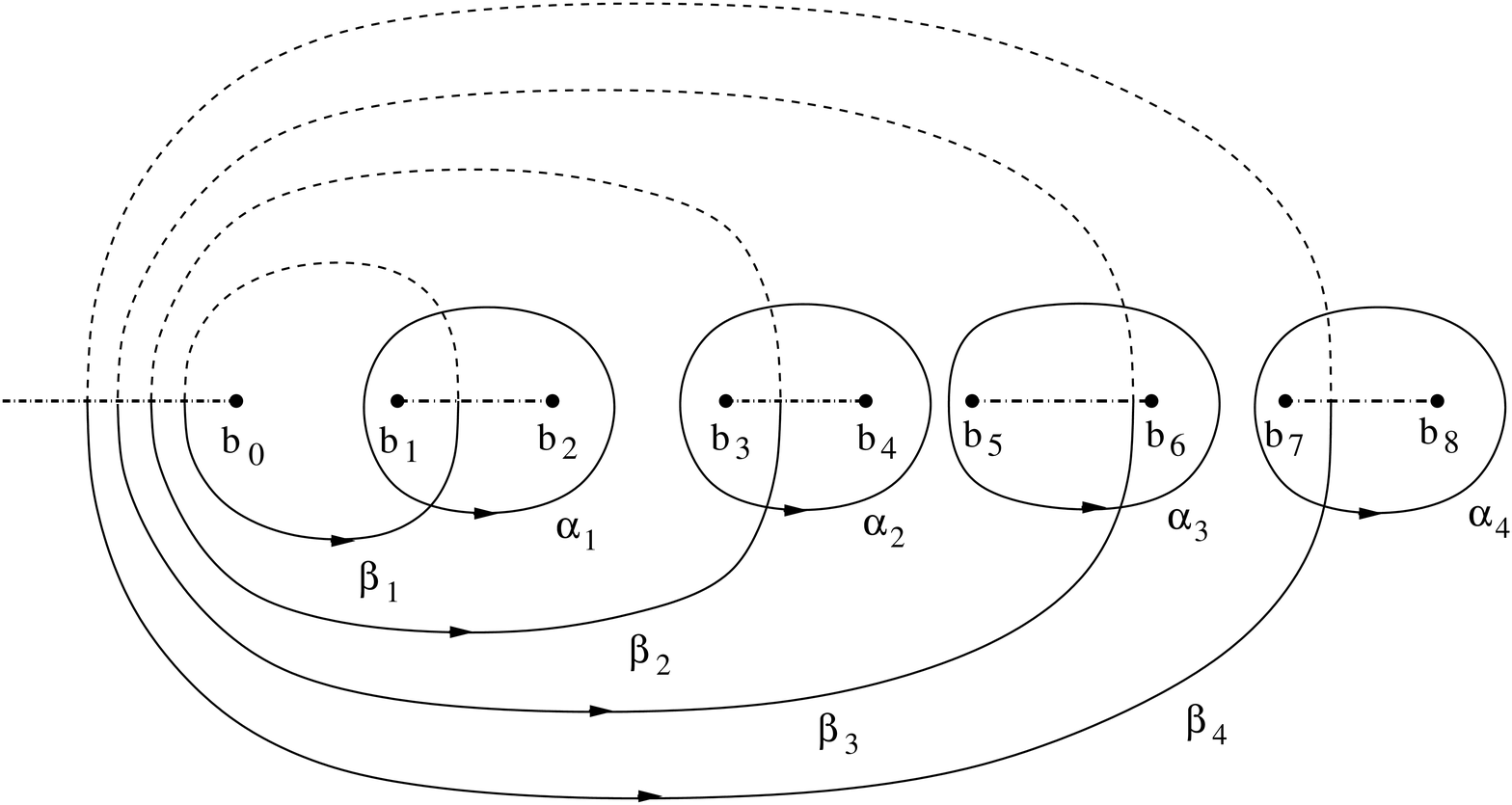}}

\centerline{Figure 1: The canonical basis of cycles.}
\end{center}

\section{The billiards on the ellipsoid  $Q$.}
The billiard map is associated to the
geodesic motion of a point on $Q$ with elastic
bounces along a set of  confocal quadrics $ Q_{d_s}$,
$s=1,\ldots,r$, $d_1<d_2<\ldots<d_r$. We are interested in real motions
only, and we assume, that all reflectors  $Q\cap Q_{d_s}$ are met at
least once. Therefore we have the following constraints to the positions
of the reflectors:
\begin{enumerate}
\item The points $d_s$ are located inside gaps: $d_s\in \bigcup
\limits_{j=1}^n ] b_{2j-1}, b_{2j} [, \quad s=1,\ldots,r$ \item
Each gap $[b_{2j-1},b_{2j}]$ contains at most 2 points  $d_s$.
\item If the gap $[b_{2j-1},b_{2j}]$  contains exactly 2 points
$d_s$, $d_{s+1}$, then the projection of the divisor point
$\la_j(t)$ is located inside the interval $[d_s,d_{s+1}]$. \item
If the gap  $[b_{2j-1},b_{2j}]$ contains exactly 1 point $d_s$,
  then the projection of the divisor point $\la_j(t)$ remains forever inside one
  of the following intervals: $[b_{2j-1},d_s]$ or $[d_s,b_{2j}]$.
\end{enumerate}

The divisor dynamics can be described in the following way. It is
guided by the Dubrovin equations (\ref{eq:dubrovin}) between
the subsequent collisions with the reflectors. If a divisor point $\la_j$
collides with reflector $d_s$, it ``jumps'' to the opposite sheet of
$\Gamma$ with respect to the hyperelliptic involution:
\begin{equation}
\tau: (\la_j,w_j)\rightarrow (\la_j,-w_j).
\end{equation}
Each point $d_s$ has two preimages in the spectral curve $\Gamma$:
$D_s^+=(d_s,+w(d_s))$, $D_s^-=(d_s,-w(d_s))$, $w(d_s)>0$. Let us
introduce the following notation:
\begin{eqnarray}
\label{eq:dsinout}
D_s^{\rm{in}} & = & \left\{\begin{array}{cc} D_s^+ \quad \mbox{if} &
    \displaystyle \frac{(\la_j-d_s)}{\prod\limits_{k\ne j}(\la_j-\la_k)}
    >0 \\
D_s^- \quad \mbox{if} &
    \displaystyle \frac{(\la_j-d_s)}{\prod\limits_{k\ne j}(\la_j-\la_k)}
    <0
  \end{array}\right.\\ \\
D_s^{\rm{out}} & = & \tau D_s^{\rm{in}}  \nonumber
\end{eqnarray}
From the Dubrovin equations (\ref{eq:dubrovin}) it follows, that the
divisor points always jump from $D_s^{\rm{in}}$ to $D_s^{\rm{out}}$.

Under the Abel--Jacobi map (\ref{AB}), the condition $\la_j=d_s$
defines two codimension one subvarieties $\Theta^{\rm{in}}_s$ and
 $\Theta^{\rm{out}}_s$ in ${\rm Jac} (\Gamma)$,
and each reflection on the $Q\cap Q_{d_s}$ corresponds to the jump
from one of these subvarieties to the other one.

Hence, for fixed constants of motion, the coordinates $x$ of
impact points on $Q\cap  Q_{d_s}$ with velocities $v$ are described by
degree $n-1$ divisors $\{P_1,\dots,P_{n-1}\}$, that is, by a point
$\varphi$ on $\Theta^{\rm{in}}_s$ or $\Theta^{\rm{out}}_s$.

Let us denote
\begin{equation} \label{qqs}
{\mathfrak q}_s =({\mathfrak q}_s^1,\dots,{\mathfrak q}_s^n)^T =
\int_{D_s^{\rm{in}}}^{D_s^{\rm{out}}} (\omega_1,\dots,\omega_n)^T \in {\mathbb
R}^{n}.
\end{equation}

Let us recall the situation in case of one reflector $d_1=d$,
$\Theta^{\rm{in}}_1=\Theta^{\rm{in}}$,
$\Theta^{\rm{out}}_1=\Theta^{\rm{out}}$,
$D^{\rm{in}}=D^{\rm{in}}_1$,  $D^{\rm{out}}=D^{\rm{out}}_1$,
${\mathfrak q}={\mathfrak q}_1 $. The billiard map is
transcendental, as it was first observed by Veselov for $n=2$
\cite{Ves2}.  An algebraic geometric description of the motion
between impacts and at the elastic bounce along $ Q\cap  Q_d$ is
given by the following proposition \cite{AF2}.

\begin{theorem} \label{main0}
The geodesic motion on $Q$ between subsequent impact points
with coordinate $x^{(N)}$ and $x^{(N+1)}$ on $Q\cap Q_d$
corresponds to the straight line uniform motion on ${\rm Jac}
(\Gamma)$ between $\Theta^{\rm{out}}$ and $\Theta^{\rm{in}}$ along the
$u_1$-direction. The point of intersection with $\Theta^{-}$ gives
the next coordinate $x^{(N+1)}$ and the ingoing velocity
$v^{(N+1)}_{\rm{in}}$. At the point $x^{(N+1)}$ the reflection
$v^{(N+1)}_{\rm{in}} \mapsto v^{(N+1)}_{\rm{out}}$ (from the ingoing to the
outgoing velocity) results in jumping back from  $\Theta^{\rm{in}}$ to
$\Theta^{\rm{out}}$ by the shift vector $\mathfrak q$, which does not
change $x^{(N+1)}$.  Then the procedure iterates.
\end{theorem}

The process is sketched in Figure 2. Here $A^{N}_{\rm{in}}$ denotes
the image through the Abel map of the divisor, associated to the
point $(x^{(N)},v^{(N)}_{\rm{in}})$, $A^{N}_{\rm{out}}$ denotes the image of
$(x^{(N)},v^{(N)}_{\rm{out}})$ respectively:

\begin{center}
{\epsfxsize=10cm \epsffile{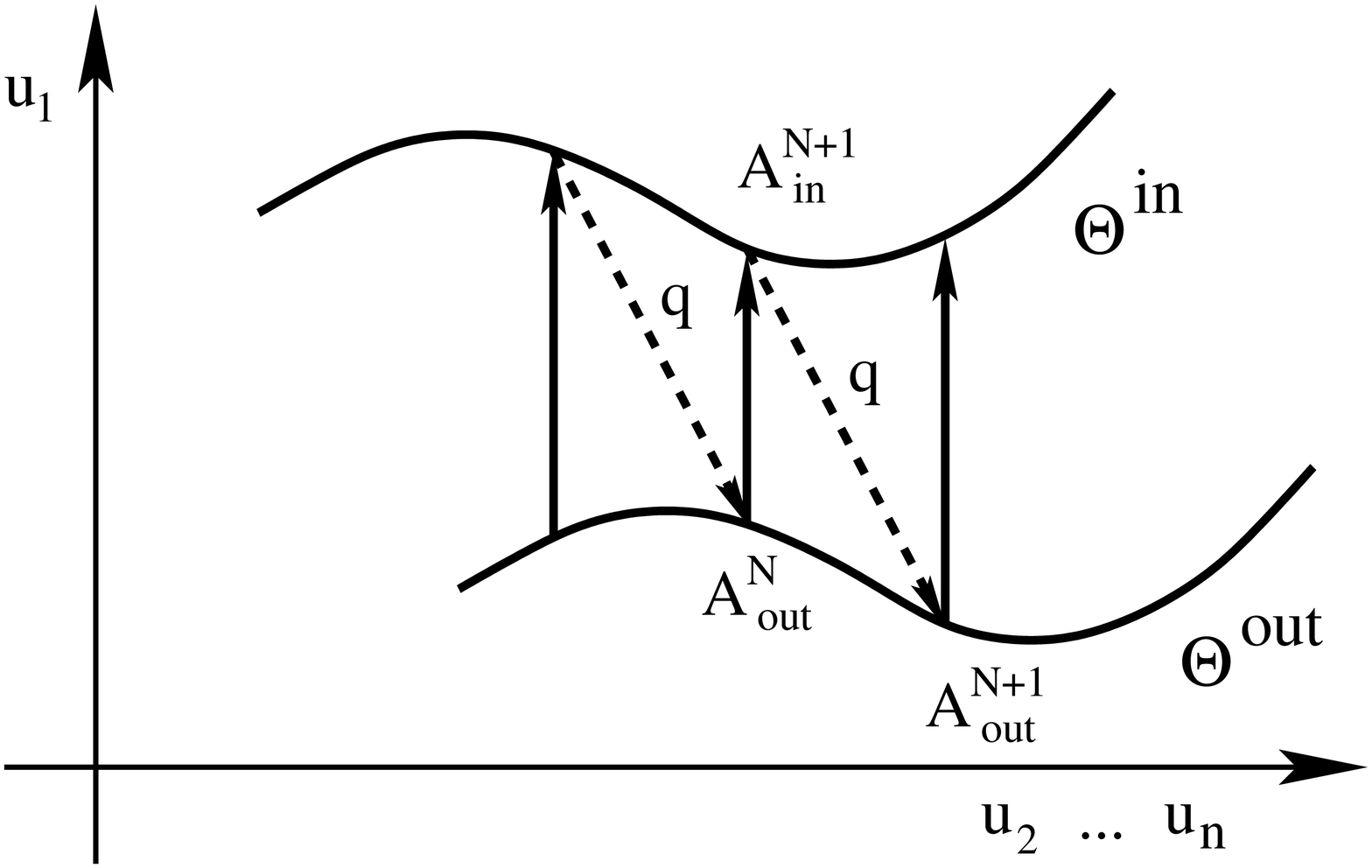}}

\centerline{Figure 2: The billiard dynamics in the Jacobian.}
\end{center}

To study the billiard dynamics it is important to choose an
appropriate canonical basis of cycles. We assume that $\Gamma$ has
the following system of cuts: an infinite cut $]-\infty,0]$ and
finite cuts over gaps $[b_{2j-1},b_{2j}]$ (all cuts are real).
We choose the $\alpha$-cycles in the standard way (see Figure 3). It
is important that they do not intersect the cuts, and
consequently, the trajectories of the divisor. If the gap
$[b_{2j-1},b_{2j}]$ does not contain reflectors, the corresponding
$\beta_j$-cycle is also chosen in the standard way, and it
intersects exactly once the cut $[b_{2j-1},b_{2j}]$. If the gap
$[b_{2j-1},b_{2j}]$ contains at least one reflector, we choose the
cycle  $\beta_j$ so, that it does not intersect the divisor
trajectory. In the case of two reflectors in the gap
$[b_{2j-1},b_{2j}]$ there are two possible choices of $\beta_j$
(see cycle $\beta_3$ in Figure 3).

\begin{center}
{\epsfxsize=13cm \epsffile{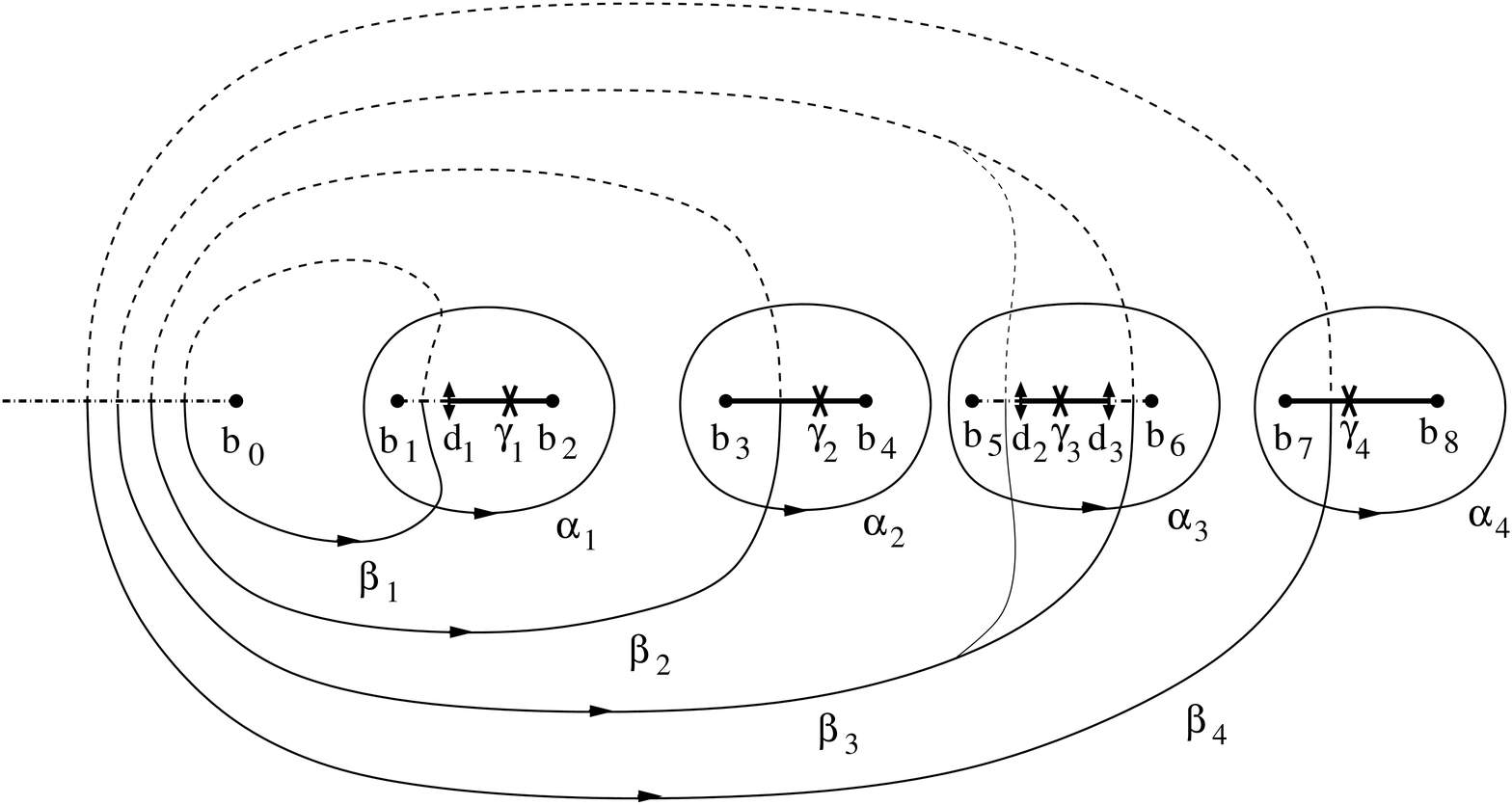}}

\centerline{Figure 3: The canonical basis of cycles in the presence of
reflectors.}
\end{center}

Let us us also introduce an auxiliary set of curves
$\tilde\alpha_1$, \ldots, $\tilde\alpha_n$. If the gap
$[b_{2j-1},b_{2j}]$ does not contain reflectors, $\tilde\alpha_j$
is the preimage of $[b_{2j-1},b_{2j}]$ with the proper orientation
and homologically $\tilde\alpha_j \sim\alpha_j$. If the gap
$[b_{2j-1},b_{2j}]$ contains reflectors, $\tilde\alpha_j$ is the
part of the preimage of $[b_{2j-1},b_{2j}]$ where the divisor
moves. The curves $\tilde\alpha_j$ inherit the orientations from
the basic cycles $\alpha_j$.

Let us define the auxiliary basis $\tilde\omega_1$,\ldots,
$\tilde\omega_n$ of holomorphic differentials with the
following {\bf non-standard} normalization:
\begin{equation}
\label{eq:ourbasis}
\int\limits_{\tilde\alpha_j} \tilde\omega_k =\delta_{jk}.
\end{equation}

\begin{proposition}
A basis $\tilde\omega_1$,\ldots, $\tilde\omega_n$ with
normalization (\ref{eq:ourbasis}) exists and is unique.
\end{proposition}

The existence and the uniqueness of such basis of holomorphic
differentials, follows immediately from the following lemma.

\begin{lemma}
\label{lemma:nondegeneration}
Let $\mathcal{G}_{(1)}$, \ldots, $\mathcal{G}_{(N)}$,  $\mathcal{G}=\cup \mathcal{G}_{(j)}$
be a set of paths in the cycles $\alpha_k$ such, that
\begin{enumerate}
\item The velocity vector for all paths  $\mathcal{G}_{(j)}$ is
  everywhere non-zero.
\item All paths $\mathcal{G}_{(j)}$ belonging to the same $\alpha_k$ have the
  same orientation with respect to the standard orientation on
  $\alpha_k$. Let us denote: $\mathcal{G}_k=\mathcal{G}\cap\alpha_k$, i.e.
 $\mathcal{G}_k$ is the union of all trajectories, lying in the cycle $a_k$.
\end{enumerate}
Then for any $k$ such, that $\mathcal{G}_k\ne\varnothing$ there exists a
holomorphic differential $\omega$ such, that
\begin{equation}
\label{eq:nondegeneration}
\int\limits_{\mathcal{G}_l} \omega =\delta_{kl}.
\end{equation}
\end{lemma}

Proof of Lemma~\ref{lemma:nondegeneration}: Let us
consider the following holomorphic differential
\begin{equation}
\label{eq:omega}
\omega = \mathcal{C} \frac{\prod\limits_{j\ne k} \left(\lambda-{\tilde \eta_j}
  \right) d\la }{w},
\end{equation}
where $\tilde \eta_j\in\alpha_j$. Differential $\omega$ has a constant
sign on $\alpha_k$, therefore
\begin{equation}
\label{eq:nondegeneration2}
\int\limits_{\mathcal{G}_k } \omega \ne  0.
\end{equation}
If $\mathcal{G}_j=\varnothing$ then
\begin{equation}
\label{eq:nondegeneration3}
I_j=\int\limits_{\mathcal{G}_j } \omega =0.
\end{equation}
If  $\mathcal{G}_j\ne\varnothing$ and $\tilde\eta_j$ lies in
the end of the interval $[b_{2j-1},b_{2j}]$ then integral $I_j$ is
nonzero. The signs of $I_j$ for $\tilde\eta_j=b_{2j-1}$ and
$\tilde\eta_j=b_{2j}$, are opposite, and they do not depend
on the position of other zeroes $\tilde\eta_i$. Therefore we
can tune the set $\tilde\eta_j$ so, that all $I_j=0$ for all $j\ne
k$. To complete the proof is it sufficient to choose a proper
normalization constant $\mathcal{C}$.
\begin{remark}
This proof uses the following topological Lemma: Let $f=(f^1(x_1,\ldots,x_n),\ldots,
f^n(x_1,\ldots,x_n))$ be a map from the $n$-dimensional cube $I^n$: $-1\le x_k \le 1$, $k=1,\ldots,n$ to
the $n$-dimensional space ${\mathbb R}^n$ with the following property: for all $k$ $f^k(\vec x) <0$
at the face $x_k=-1$ and $f^k(\vec x) >0$ at the face $x_k=1$. Then the preimage of the
origin $\vec x=(0,\ldots,0)$ is non-empty.

The proof of the Lemma is based on standard topological arguments and
we do not present it here.
\end{remark}
\begin{corollary}
\label{corollary:nondegenerate}
Let $\hat\omega_1$,\ldots, $\hat\omega_n$ be the {\bf standard} basis
of holomorphic differentials
\begin{equation}
\label{eq:srdbasis}
\int\limits_{\alpha_j} \hat\omega_k =\delta_{jk},
\end{equation}
$\mathcal{J}$ be a subset of the set $\{1,2,\ldots,n\}$ such, that for
any $k\in\mathcal{J}$  $\mathcal{G}_k\ne\varnothing$.
Then the matrix
\begin{equation}
\mathcal{A}_{kl} =\int\limits_{\mathcal{G}_k} \hat\omega_l, \ \ k,l\in \mathcal{J}
\end{equation}
is non-degenerate.
\end{corollary}
To prove this Corollary it is sufficient to assume, that for all $k$
such, that $1\le k\le n$, $k\not\in\mathcal{J}$ $\mathcal{G}_k=\alpha_k$ and
to apply Lemma~\ref{lemma:nondegeneration}. The differential obtained
has zero integral over all contours $\alpha_k$,  $k\not\in\mathcal{J}$,
therefore it is a linear combination of $\hat\omega_l$, $l\in\mathcal{J}$.

\paragraph{Periodic billiard trajectories}

First of all, let us recall, that there are two different
descriptions of periodicity for the billiard motion. If the
billiard trajectory is closed, it implies, that the divisor motion
is periodic with the same period. From the periodicity of the
divisor motion it follows, that the motion on the ellipsoid is
also periodic, but the period may be greater, since the same
divisor corresponds to $2^{n+1}$ different points in the phase
space assuming all constants of motion to be fixed. These points
can be obtained one from another by the reflections
$x_j\rightarrow-x_j$. We see, that each periodic divisor motion
corresponds to $2^{n+1}$ different periodic trajectories.

In the following we study necessary and sufficient conditions for the
periodicity of the divisor motion.

Let us recall the necessary and sufficient condition on the
spectral data to generate periodic billiard trajectories. In our
setting for arbitrary number of reflectors we have:
\begin{lemma}
The billiard trajectory is periodic if and only if there exist a
collection of integers $m_1$,\ldots,$m_n$ such that
\begin{equation}
\label{eq:periodicity1}
\sum\limits_{j=1}^n m_j \int\limits_{\tilde\alpha_j} \omega_k =0,
\quad k=2,\ldots,n.
\end{equation}
In such a case the length of the billiard trajectory in the
rescaled variable $\sigma$ is given by
\begin{equation}
\label{eq:periodicity2}
\sum\limits_{j=1}^n m_j \int\limits_{\tilde\alpha_j} \omega_1 = 2 L.
\end{equation}
\end{lemma}
Here $\omega_k$ is the basis of holomorphic differentials
(\ref{basis}) normalized by the asymptotics near the Weierstrass point
$\la=0$.

The above conditions were also given in \cite{DragRad} in an
equivalent formulation and are in general transcendental in the
parameters of the problem (the square semiaxes, the constants of
the motion and the position of reflectors). In the case of one
reflector, in \cite{AF2} the authors gave a sufficient condition
so that the periodicity condition for the billiard trajectory is
algebraic in the parameters.

To apply the method of isoperiodic deformations we need an alternative
characterization of the periodic spectral data. The remaining part of
this Section is dedicated to this alternative characterization in
terms of the so-called generalized quasimomentum $d\tilde p$.

First of all we assume that our billiard motion is periodic, and we
construct the generalized quasimomentum, satisfying the properties 1-9 in
Lemma~\ref{lemma:xi2}. Then we show that a unique generalized
quasimomentum satisfying properties 1-7  in Lemma~\ref{lemma:xi2}
exists for general quasiperiodic solutions. Finally we check, that the
motion is periodic if and only if the properties 8-9 are fulfilled.

Assume, that the trajectory is periodic with the period $T$.
Denote the impact points by $t_1$,\ldots,$t_M$,
$0<t_1<\ldots<t_M<T$. Consider the following function:
\begin{equation}
\label{eq:xi}
\Xi(\gamma)= \frac{\Psi(\gamma,t_1-0)}{\Psi(\gamma,0)}
\frac{\Psi(\gamma,t_2-0)}{\Psi(\gamma,t_1+0)}\cdots
\frac{\Psi(\gamma,T)}{\Psi(\gamma,t_M+0)}
\end{equation}
\begin{lemma}
The function $\Xi(\gamma)$ has the following analytic properties:
\begin{enumerate}
\item $\Xi(\gamma)$ is holomorphic in $\Gamma$ outside the points
  $\lambda=0$ and impact points $D^{\pm}_s$.
\item
\begin{equation}
\Xi(\gamma)= 1 +\frac{iT}{\sqrt{\la}}+\ldots , \quad \mbox{as} \quad
\lambda\rightarrow\infty
\end{equation}
\item There exists a constant $L$ such, that
\begin{equation}
\label{eq:bloch_mult}
\exp{\left(-\frac{iL}{\sqrt{\la}}\right)}\Xi(\gamma)
\end{equation}
is regular near $\la=0$.
\item  $\Xi(\gamma)$ has a pole of order ${\tt m}_s$ at the point
  $D_s^{\rm{out}}$ and a zero of order  ${\tt m}_s$ at the point
  $D_s^{\rm{in}}$ where ${\tt m}_s$ denotes the number of impacts to the
  reflector $Q\cap Q_s$.
\end{enumerate}
\end{lemma}

Let us introduce generalized quasimomentum $\tilde p$ for the
billiard motion. We would like to stress, that the generalized
quasimomentum {\bf does not coincide} with the ``true''
quasimomentum for the non-smooth function $u(t)$ describing the
billiard motion. The ``true'' quasimomentum is defined on an
infinite-genus Riemann surface.

By definition:
\begin{equation}
\label{eq:dptilde}
\tilde p = \frac{1}{iT} \ln\left(\Xi(\gamma) \right)
\end{equation}
The function $\tilde p(\gamma)$ is multivalued in $\Gamma$ but its
differential  $d\tilde p$ is well-defined.
\begin{lemma}
\label{lemma:xi1} The function $\tilde p(\gamma)$ is a
single-valued function on
$\Gamma^*=\Gamma\backslash\left(\bigcup\limits_{j=1}^{n}\tilde\alpha_j\right)$.
In other words $\Gamma^*$ is obtained from $\Gamma$ by removing
the divisor trajectories.
\end{lemma}
The proof of the Lemma follows immediately from the formula:
\begin{equation}
\label{eq:ptilde}
\tilde p= \frac{1}{iT}\left[\int\limits_{0}^{t_1}
\frac{\dot\Psi(\gamma,t)}{\Psi(\gamma,t)} dt +
\int\limits_{t_1}^{t_2}
\frac{\dot\Psi(\gamma,t)}{\Psi(\gamma,t)} dt + \ldots +
\int\limits_{t_M}^{T}
\frac{\dot\Psi(\gamma,t)}{\Psi(\gamma,t)} dt \right]
\end{equation}
The formula is well-defined if the denominator does not vanish.
\begin{lemma}
\label{lemma:gap} Assume, that the gap $[b_{2j-1},b_{2j}]$
contains at least one reflector, and that $\gamma\in\Gamma$ lies
inside the gap but outside the divisor trajectory. Then
$\Re(\Xi(\gamma))=0$.
\end{lemma}
The proof follows immediately from the formula (\ref{eq:ptilde}).

Now we are ready to formulate the properties of the generalized
quasimomentum.
\begin{lemma}
\label{lemma:xi2}
\begin{enumerate}
\item\label{property:1}
 $d\tilde p$ is a meromorphic differential in $\Gamma$.
\item $d\tilde p$ is holomorphic outside the points $\la=0$,
  $D_s^{\rm{in}}$, $D_s^{\rm{out}}$.
\item $d\tilde p$ has first-order poles at the points
$D_s^{\rm{in}}$, $D_s^{\rm{out}}$ and
\begin{equation}
\res[D_s^{\rm{in}}] d\tilde p =-\res[D_s^{\rm{out}}] d\tilde p
\end{equation}
If we have two reflectors  $d_s$, $d_{s+1}$ in the same gap, then
\begin{equation}
\res[D_s^{\rm{in}}]d\tilde p =\res[D_{s+1}^{\rm{in}}] d\tilde p.
\end{equation}
\item $d\tilde p$ has a second-order pole with zero residue at
  $\la=0$.
\item
\begin{equation}
\label{eq:property:5}
d\tilde p =
d\left(\frac{1}{\sqrt{\la}}\right)+O\left(\frac{1}{\la^2}\right)d\la,
\quad \mbox{as} \; \lambda\to \infty.
\end{equation}
\item\label{property:6}
\begin{equation}
\oint\limits_{\alpha_j} d\tilde p=0, \quad j=1,\ldots,n.
\end{equation}
\item
\label{property:7}
\begin{equation}
\label{eq:property:7}
\oint\limits_{\beta_j} d\tilde p=0, \quad \mbox{if the gap} \ \
[b_{2j-1},b_{2j}] \ \ \mbox{contains at least one reflector}.
\end{equation}
Let us recall that for gaps with reflectors the cycles $\beta_j$
do not intersect the divisor trajectories. \item\label{property:8}
\begin{equation}
\res[D_s^{\rm{in}}]d\tilde p =\frac{{\tt m}_s}{iT}, \quad {\tt m}_s\in{\mathbb N}
\end{equation}
where ${\tt m}_s$ denotes the number of impacts to the obstacle $Q\cap Q_s$
for $0\le t<T$.
\item\label{property:9}
\begin{equation}
\oint\limits_{\beta_j} d\tilde p=\frac{2\pi n_j}{T}, \ \
n_j\in{\mathbb Z} \quad \mbox{if the gap} \ \
[b_{2j-1},b_{2j}] \ \ \mbox{contains no reflectors}.
\end{equation}
\end{enumerate}
\end{lemma}
Properties~\ref{property:6},~\ref{property:7} follow immediately
from Lemmata~\ref{lemma:xi1},\ref{lemma:gap}, and the fact that
these cycles do not intersect the divisor trajectories is
essential. All other properties are completely standard.

\begin{remark}
Differential $d\tilde p$ is even with respect to the transposition of
the sheets of $\Gamma$. Therefore the second term in (\ref{eq:property:5})
can be estimated more accurately:
\begin{equation}
\label{eq:property:5bis}
d\tilde p =
d\left(\frac{1}{\sqrt{\la}}\right)+O\left(\frac{1}{\la^{5/2}}\right)d\la,
\quad \mbox{as} \; \lambda\to \infty.
\end{equation}
At $\la=0$ one has the following expansion:
\begin{equation}
\label{eq:property:4bis}
d\tilde p =
\frac{L}{T} \ d\left(\frac{1}{\sqrt{\la}}\right) + O(1) d(\sqrt{\la}),
\end{equation}
where $L$ is the same as in (\ref{eq:bloch_mult}).
\end{remark}

Let us have a generic spectral curve $\Gamma$ with a set of
reflectors $d_s$ (with the reality conditions, formulated above).
\begin{lemma}
\label{lemma:qusimomentum}
There exists an unique differential with the properties
\ref{property:1}-\ref{property:7}.
\end{lemma}
This differential plays the
role of the generalized quasimomentum differential for generic
billiard spectral data, and we shall denote it with the same
symbol $d\tilde p$.

Proof of Lemma~\ref{lemma:qusimomentum}. Let us construct $d\tilde p$
as the following linear combination:
\begin{equation}
\label{eq:dtildep-generic}
d\tilde p =\mathcal{C}dp', \ \ \mbox{where} \ \ dp' = d\hat p + \sum\limits_{k\in \mathcal{J}}
\mathcal{D}_k \Omega_{(k)},
\end{equation}
$\mathcal{J}$ denotes the set of all $\alpha$-cycles, containing
reflectors, $d\hat p$ is the Abelian differential of the second kind with a
second-order pole at the point $\la=0$ such, that
\begin{equation}
d\hat p = d\left(\frac{1}{\sqrt{\la}}\right)+ \ldots \ \ \mbox{as} \ \
\la\rightarrow 0, \qquad \oint\limits_{\alpha_j} d\hat p=0, \quad j=1,\ldots,n,
\end{equation}
$\Omega_{(k)}$ are the Abelian differentials of the third kind with
first-order poles at the reflector points in the  $\alpha_k$,
\begin{equation}
\res[D_s^{(in)}] \Omega_{(k)} = +1, \ \ \res[D_s^{(out)}]\Omega_{(k)}=-1, \ \
D_s^{(in)}, D_s^{(out)} \in \alpha_k,
\end{equation}
\begin{equation}
\oint\limits_{\alpha_j} \Omega_{(k)} =0, \quad j=1,\ldots,n.
\end{equation}
From the Riemann bilinear relations it follows, that
\begin{equation}
{\tt B}_{lk}= \oint\limits_{\beta_l} \Omega_{(k)} =
-2\pi i  \int\limits_{\tilde\alpha_k}\hat\omega_l.
\end{equation}
By Corollary~\ref{corollary:nondegenerate} the matrix ${\tt B}_{lk}$,
$l,k\in\mathcal{J}$ is non-degenerate, therefore conditions
(\ref{eq:property:7}) defines the constants $\mathcal{D}_k$ uniquely.

Let us consider the divisor dynamics with reflections in $\Gamma$,
written in terms of the KdV time $\sigma$ (see (\ref{tau-1})). Then outside the
divisor trajectories we have:
\begin{equation}
dp'=\lim\limits_{L\rightarrow \infty} \frac{1}{iL} \int\limits_0^L
\frac{\Psi'(\gamma,\sigma)}{\Psi(\gamma,\sigma)} d\sigma
\end{equation}
For the logarithmic derivative of the wave function we have the following formula
\begin{equation}
\label{eq:dPsiprime}
\frac{\Psi'(\gamma,\sigma)}{\Psi(\gamma,\sigma)} =
\frac
{\sqrt{\prod\limits_{j=1}^{2n} b_j}}
{\prod\limits_{j=1}^n \lambda_j(\sigma)}
\cdot\frac{w+\lambda{R}_{n-1}(\lambda)}{
\lambda(\lambda-\lambda_1(\sigma))\ldots(\lambda-\lambda_n(\sigma))},
\end{equation}
where ${R}_{n-1}(\lambda)$ denotes some polynomial of degree
$n-1$. For $\lambda\rightarrow\infty$ we have
\begin{equation}
\label{eq:dPsiprime2}
\frac{\Psi'(\gamma,\sigma)}{\Psi(\gamma,\sigma)} \sim
\frac{\sqrt{\prod\limits_{j=1}^{2n} b_j}}{\prod\limits_{j=1}^n \lambda_j(\sigma)}
\cdot\frac{i}{\sqrt{}\lambda},
\end{equation}
therefore the leading term at $\infty$ does not vanish and has a
constant sign. As a corollary the leading term of $dp'$ at $\infty$
does not vanish, and the constant $\mathcal{C}$ is defined by 
(\ref{eq:property:5}). 

Lemma~\ref{lemma:qusimomentum} is proved.

Let us formulate the sufficient conditions for the periodicity of
the billiard trajectory.
\begin{proposition}
\label{proposition:period}
Assume that the quasimomentum differential $d\tilde p$, defined by the
properties  \ref{property:1}-\ref{property:7} satisfy also
properties \ref{property:8}, \ref{property:9} for some $T$.

Then the
corresponding divisor billiard motion is periodic and with the period $T$ in
the natural parameter, and there exists an integer $K$ such, that the
corresponding billiard trajectory on the ellipsoid is periodic with
the period $KT$.

\end{proposition}

Let us consider the divisor dynamics with reflections in $\Gamma$,
written in terms of the KdV time $\sigma$ (see (\ref{tau-1})).
Denote by $\tilde A(t)$ the Abel transform of the divisor expanded
by the auxiliary basis $\tilde\omega_1$,\ldots,  $\tilde\omega_n$.
From a standard calculation based on the Riemann bilinear
identities it follows that
$$
\frac{d\tilde A_j(\sigma)}{d\sigma}=
\left\{\begin{array}{ll}  \frac{{\tt m}_j}{L} & \mbox{if the gap} \ \
[b_{2j-1},b_{2j}] \ \mbox{contains reflector points} \\ \\
\frac{n_j}{L} & \mbox{if the gap} \ \ [b_{2j-1},b_{2j}] \
\mbox{contains no reflectors.}
\end{array} \right. ,
$$
where $L$ is the same as in (\ref{eq:bloch_mult}), (\ref{eq:property:4bis}).

Let us denote by $\mathcal{G}_1$ the union of all trajectories
generated by the divisor motion for $0\le\sigma\le L$ and by
$\mathcal{G}_0$ the following collection of oriented curves:
$$
\mathcal{G}_0 = \left\{ \bigcup\limits_{\mbox{all gaps containing reflectors}} {\tt
  m}_j \tilde\alpha_j \right\} \bigcup \left\{
\bigcup\limits_{\mbox{all gaps without reflectors}} n_j
\tilde\alpha_j \right\}.
$$
From the Dubrovin equations (\ref{eq:dubrovin}) it follows that the
direction of motion in each gap is constant. Therefore
the difference between  $\mathcal{G}_1$ and  $\mathcal{G}_0$ satisfies the
conditions of Lemma~\ref{lemma:nondegeneration}.
The Abel transform for this difference is zero, therefore these two sets of
paths coincide. It implies the periodicity of the divisor
motion with the period $L$ in the variable $\sigma$. Following the procedure
described above one can naturally introduce an analog of the generalized
quasimomentum with respect to the variable $\sigma$. Let us denote it
by $dp'$. It coincides with  $d\tilde p$ up to a normalization
constant, and it is easy to check, that $dp'=\frac{T}{L}d\tilde p$,
therefore the divisor motion is also periodic in $t$ with the period $T$.
\begin{remark}
Let us show, that both of the periodicity conditions formulated above are equivalent.
Equation (\ref{eq:periodicity1}) defines a collection of numbers $m_j$ uniquely up
to a constant multiplier. From the bilinear relations we see that they are
proportional to either residues or $\beta$-periods of the generalized quasimomentum
differential  $d\tilde p$ whether or not the corresponding gap contains reflectors.
Proposition~\ref{proposition:period} means exactly that in the periodic case one can choose
the constant multiplier so that all  $m_j$ are integer numbers.
\end{remark}

\section{Isoperiodic deformations for the billiard motion}

Denote the zeroes of the quasimomentum differential $d\tilde p$ by
$\eta_1$,\ldots,$\eta_{n+r}$. Then the quasimomentum differential
takes the form:
\begin{equation}
\label{eq:ptildeqp} d\tilde p = -\frac{1}{2}
\frac{\prod\limits_{j=1}^{n+r}(\la-\eta_j)} {\la
\prod\limits_{s=1}^{r}(\la-d_s) \sqrt{\la
\prod\limits_{j=1}^{2n}(\la-b_j)}}d\la
\end{equation}

By analogy with \cite{Grinevich} we shall use the following
statement:
\begin{lemma}
Assume, that we have a periodic spectral data for the billiard
motions and a deformation of these data such, that
\begin{equation}
\label{eq:deform1}
\left.\frac{\partial \ d\tilde p}{\partial\zeta}\right|_{\la={\rm
    const}} = d P(\gamma),
\end{equation}
or, equivalently,
\begin{equation}
\label{eq:deform2}
\left.\frac{\partial \tilde p}{\partial\zeta}\right|_{\la={\rm const}} = P(\gamma),
\end{equation}
where $P(\gamma)$ is a meromorphic function in $\Gamma$. Then this
deformation preserves the $T$-periodicity of the divisor motion
associated to the billiard dynamics.
\end{lemma}
In our situation the spectral curve depends on the variable
$\zeta$. While calculating the partial derivatives with respect to
$\zeta$ we assume the projection of the point $\gamma$ to the
$\la$-plane to remain constant. We remark that this rule
generates singularities at the moving branch points.

\begin{remark}
The function $\tilde p$ in the left-hand side of (\ref{eq:deform2}) is
multivalued. The condition that its derivative is single-valued is
equivalent to the isoperiodicity condition (see \cite{Grinevich}).
\end{remark}

The proof of the Lemma is  rather straightforward. From the equation
(\ref{eq:deform1}) it follows that all periods of $d\tilde p$ as well
as the residues at the points $D_s^{\rm{in}}$,  $D_s^{\rm{out}}$ are
constant, therefore the properties  \ref{property:8}, \ref{property:9}
are satisfied identically.
\begin{lemma}
\label{lemma-dp2} Assume, that we have a deformation of the
periodic spectral data satisfying (\ref{eq:deform1}). Then
$P(\gamma)$ has the following analytic properties:
\begin{enumerate}
\item\label{prop:P1}
 $P(\gamma)$ has 1-st order poles at the points $\la=0$ and
$D_s^{\rm{in}}$,  $D_s^{\rm{out}}$.
\item\label{prop:P2}
$P(\gamma)$ has 1-st order poles at the finite ramification
  points $b_1$,\ldots,$b_{2n}$.
\item\label{prop:P3} $P(\gamma)$ has a third-order zero at
$\la=\infty$. (Here we use
  that $d\tilde p$ is even with respect to the hyperelliptic
  involution $\tau$).
\item\label{prop:P4} $P(\gamma)$ is odd with respect to the
hyperelliptic involution $\tau$: $P(\tau\gamma)=-P(\gamma)$.
\end{enumerate}
\end{lemma}
The proof again is rather straightforward. We prove now that the
opposite is also true.
\begin{lemma}
\label{lemma-dp3}
Assume that $P(\gamma)$ is an arbitrary meromorphic function
satisfying the properties \ref{prop:P1}-\ref{prop:P3} from
Lemma~\ref{lemma-dp2}. Then it generates an infinitesimal isoperiodic
deformation.
\end{lemma}
This lemma generalizes the analogous statement from
\cite{Grinevich}. The idea how to construct this deformation is
very similar to the arguments of the paper \cite{FFM}.

First of all let us define the dynamics of the branch points. Consider
the expansions of $\tilde p$, $P$ near the point $b_j$:
\begin{equation}
\label{eq:deform3}
\tilde p = {\tilde p}_{0,b_j}(\zeta)+
{\tilde p}_{1,{b_j}}(\zeta)\sqrt{\la-b_j(\zeta)}+\ldots \quad
d \tilde p=\left[
\frac{{\tilde p}_{1,{b_j}}(\zeta) }{2\sqrt{\la-b_j(\zeta)}}
+O(1)\right] d\la
\end{equation}
\begin{equation}
\label{eq:deform4}
P = \frac{{P}_{-1,b_j}(\zeta)}{\sqrt{\la-b_j(\zeta)}}+\ldots
\end{equation}
\begin{equation}
\label{eq:deform5}
\left.\frac{\partial\tilde p}{\partial\zeta}\right|_{\la={\rm const}} =
 -\frac{1}{2\sqrt{\la-b_j(\zeta)}}
{\tilde p}_{1,{b_j}}(\zeta) \frac{\partial b_j(\zeta)}{\partial\zeta} +\ldots
\end{equation}
Comparing (\ref{eq:deform4}) with (\ref{eq:deform5}) we obtain:
\begin{equation}
\label{eq:deform6}
\frac{\partial b_j(\zeta)} {\partial\zeta}=-2\frac{{P}_{0,d_s}(\zeta)}
{{\tilde p}_{1,{b_j}}(\zeta)}
\end{equation}
Let us define now the dynamics of the reflectors $d_s$. Near the
points $D_s^{\rm{in}}$,  $D_s^{\rm{out}}$ we have respectively:
\begin{equation}
\label{eq:deform7}
\tilde p= \pm \frac{{\tt m}_s}{iT} \ln(\la-d_s(\zeta))+\ldots \quad
d\tilde p= \pm \frac{{\tt m}_s}{iT(\la-d_s(\zeta))}d\la +\ldots
\end{equation}
\begin{equation}
\label{eq:deform8}
P = \pm \frac{{P}_{-1,d_s}(\zeta)}{(\la-d_s(\zeta))}+\ldots
\end{equation}
\begin{equation}
\label{eq:deform9}
\left.\frac{\partial\tilde p} {\partial\zeta}\right|_{\la={\rm const}}
= \mp\frac{{\tt
    m}_s} {iT (\la-d_s(\zeta))} \frac{\partial d_s(\zeta)}{\partial \zeta}
 +\ldots
\end{equation}
Comparing (\ref{eq:deform8}) with (\ref{eq:deform9}) we obtain:
\begin{equation}
\label{eq:deform10}
\frac{\partial d_s(\zeta)} {\partial\zeta}=-\frac{iT{P}_{-1,d_s}(\zeta)}
{{\tt m}_s}
\end{equation}
Consider the deformation defined by (\ref{eq:deform6}),
(\ref{eq:deform10}). Then equations (\ref{eq:deform1}),
(\ref{eq:deform2}) define a deformation of $\tilde p(\gamma)$
as well as a deformation of $d \tilde p$ such, that the deformed
function  $d \tilde p$ is defined on the new spectral curve and
satisfies all properties of Lemma~\ref{lemma:xi2}. Therefore the
deformed spectral data also generate periodic divisor motions with
the same number of impacts.

By analogy with \cite{Grinevich} it is also convenient to derive
the dynamics for the zeroes of $d \tilde p$. In the neighborhood
of $\eta_j$ we have
\begin{equation}
\label{eq:deform11}
d\tilde p= \pm {\tilde p}_{1,\eta_j} (\zeta)(\la-\eta_j(\zeta))d\la +\ldots
\end{equation}
\begin{equation}
\label{eq:deform12}
P = \pm \left[ {P}_{0,\eta_j}(\zeta)+
  {P}_{1,\eta_j}(\zeta)(\la-\eta_j(\zeta)) \right]+\ldots, \quad
dP =  \pm {P}_{1,\eta_j}(\zeta)d\la +\ldots
\end{equation}
\begin{equation}
\label{eq:deform13}
\left.\frac{\partial\ d\tilde p} {\partial\zeta}\right|_{\la={\rm const}}
= \mp \left[{\tilde p}_{1,\eta_j} (\zeta) \frac{\partial\eta_j(\zeta }
{\partial\zeta}  \right]d\la +\ldots
\end{equation}
\begin{equation}
\label{eq:deform14}
\frac{\partial \eta_j(\zeta)} {\partial\zeta}=-\frac{P_{1,\eta_j}(\zeta)}
{ {\tilde p}_{1,\eta_j} (\zeta) }
\end{equation}
We have shown that we have a one-to-one correspondence between the
period preserving deformations and meromorphic functions with
special analytic properties. By analogy with \cite{Grinevich} let
as introduce a convenient basis in the space of such meromorphic
functions. Let us define:
\begin{equation}
\label{eq:basis1}
P_j(\gamma) = - \mathcal{C}_j \frac{\la d\tilde p(\gamma)} {(\la-\eta_j)
  d\la}, \quad j=1,\ldots,n+r.
\end{equation}
In the paper by Marchenko and Ostrovski \cite{Marchenko} the following
coordinates were used on the space of spectral curves associated with
periodic potentials:
\begin{equation}
\label{eq:coord1}
\zeta_j = -i p(\eta_j), \quad j=1,\ldots,g,
\end{equation}
where $\eta_j$ are the zeroes of the quasimomentum differential.
It is convenient to use these coordinates in the method of
isoperiodic deformation. In particular, the corresponding flows
commute at regular points for these flows (all zeroes $\eta_j$ are
pairwise distinct).

This choice of local coordinates correspond to the following
choice of the factors $\mathcal{C}_j$:
\begin{equation}
P_j(\eta_j)=i.
\end{equation}
We use here, that
\begin{equation}
\label{eq:deform15}
\left.\frac{\partial \tilde p}{\partial\zeta}\right|_{\la={\rm const},
\gamma=\eta_j} = \frac{\partial \tilde p(\eta_j)} {\partial\zeta}
\end{equation}
at the zeroes of the generalized quasimomentum differential $d\tilde p$.
Therefore we obtain
\begin{equation}
\label{eq:deform15_1}
\mathcal{C}_j^2=2\frac{-\eta_j\prod\limits_{k=1}^{2n}(\eta_j-b_k)
\prod\limits_{s=1}^{r}(\eta_j-d_s)^2}{\prod\limits_{k\ne j}(\eta_j-\eta_k)^2 }
\end{equation}
It is convenient to assume $\mathcal{C}_j>0$ for real zeros of
quasimomentum differential.

Calculating the leading terms of $P_j(\gamma)$ at the branch points,
reflectors and zeroes different from $\eta_j$ we obtain:
\begin{eqnarray}
\label{eq:main_deforms}
\frac{\partial b_k}{\partial \zeta_j} & = & \mathcal{C}_j
\frac{b_k}{b_k-\eta_j} \nonumber \\
\frac{\partial d_s}{\partial \zeta_j} & = & \mathcal{C}_j
\frac{d_s}{d_s-\eta_j} \nonumber \\
\frac{\partial \eta_k}{\partial \zeta_j} & = & \mathcal{C}_j
\frac{\eta_k}{\eta_k-\eta_j}, \quad k\ne j\\
\frac{\partial \eta_j}{\partial \zeta_j} & = & \mathcal{C}_j \eta_j
\left[\sum\limits_{k \ne j} \frac{1}{\eta_j-\eta_k}-\frac{1}{2\eta_j}
-\frac{1}{2}\sum\limits_{k=1}^{2n} \frac{1}{\eta_j-b_k}
- \sum\limits_{s=1}^{r} \frac{1}{\eta_j-d_s}
\right]. \nonumber
\end{eqnarray}

Summarizing all these results we obtain:

\begin{theorem}
Let us consider a generic set of spectral data corresponding to a
periodic billiard motion (generic means that all
zeroes of $d\tilde p$ are pairwise distinct). Then equations
(\ref{eq:main_deforms}) with the normalization (\ref{eq:deform15_1})
locally define a set of $n+r$ commuting flows on the spectral data, preserving
the $T$-periodicity of the billiard motion as well as the number of
impacts during a period.
\end{theorem}
\begin{remark}
In the billiard motion it is critical to respect the reality
conditions. To keep the reality of the spectral data it is
sufficient to assume, that the times $\zeta_j$  corresponding to real
zeroes $\eta_j$ of $d\tilde p$ are real, the times corresponding
to complex conjugate pairs of zeroes $\eta_k$,
$\eta_l=\overline{\eta_k}$ are complex conjugate: $\zeta_l=\overline{\zeta_k}$.
\end{remark}

\section{Appendix. Some details of numeric check.}
To illustrate the approach developed above, some numerical
simulation were made. For computational purposes it is more
convenient to use the spectral parameter $E=1/\la$ instead of
$\la$. Equations (\ref{eq:main_deforms}) were implemented using
the standard 5-th order Runge-Kutta method. The authors used the
integrator, developed by Hairer, N{\o}rsett and Wanner \cite{HNW},
this free software package in available on the Ernst Hairer's web
site \verb+http://www.unige.ch/~hairer/+. The program was
developed as a modification of the program used in
\cite{Grinevich} to check isoperiodic approach to the KdV
equation.

The novelty of the problem was, that the calculation of the
starting point for the billiard problem required some additional
calculations. Let us discuss the search of the starting point.

It is convenient to calculate the generalized quasimomentum
differential for a small periodic perturbation of the constant
potential $u(x)=1$. We assume the period $T=1$. In our experiments
we assumed that we have one reflector in the last gap
$[b_{2n-1},b_{2n}]$. In our program the branch points are
enumerate in the inverse order $E_1=1/b_{2n}$,\ldots,
$E_{2n}=1/b_1$. For numerical reasons it is critical to make the
gap, containing the reflector, a little bigger than the other
ones. We also assume, that at the starting point the reflection
wall lies in the center of the distinguished gap. The starting
lengths are: $E_2-E_1\sim 10^{-6}$ and $E_{2k}-E_{2k-1}\sim
10^{-9}$, $k>1$.

First of all, let us calculate the positions of the point $r$ (the
center of the gap), and the zeroes $A_0=1/\eta_0$, $A_1=1/\eta_1$
(up to some higher order corrections).

Let
\begin{equation}
E_1 = {\tt C}-{\tt A}, \ \ E_2={\tt C}+{\tt A}, \ \
1/d={\tt R}={\tt C},\ \ A_0 = {\tt C}+\tilde A_0, \ \
A_1 = {\tt C}+\tilde A_1, \ \ {\tt A}\ll 1.
\end{equation}
Assuming all other gaps to have zero length (genus 1 approximation) we have
$$
d\tilde\omega_1 = \varkappa
\frac{1}{\sqrt{E(E-E_1)(E-E_2)}}dE
$$
$$
{\tt C}\sim \frac{\pi^2}{4}
$$
Let us denote
$$
E={\tt Z}+{\tt C}
$$
The constant $\varkappa$ is defined from the condition
$$
\int\limits_{{\tt Z}=0}^{{\tt Z}={\tt A}} \tilde\omega_1 = \frac{1}{2}.
$$
Expanding $d\tilde\omega_1$ near ${\tt Z}=0$ we obtain:
$$
d\tilde\omega_1=\varkappa\frac{1}{\sqrt{{\tt C}+{\tt Z}}}
\frac{d{\tt Z}}{\sqrt{{\tt Z}^2-{\tt A}^2}}=
\frac{\varkappa}{\sqrt{{\tt C}}}\left(1-\frac{{\tt Z}}{2{\tt C}}
+O({\tt Z}^2)\right)\frac{d{\tt Z}}{\sqrt{{\tt Z}^2-{\tt A}^2}}
$$
Using the table integrals
$$
\int\limits_{{\tt Z}=0}^{{\tt Z}={\tt A}}
\frac{d{\tt Z}}{\sqrt{{\tt Z}^2-{\tt A}^2}} = \frac{\pi i}{2}
$$
$$
\int\limits_{{\tt Z}=0}^{{\tt Z}={\tt A}}
\frac{{\tt Z}d{\tt Z}}{\sqrt{{\tt Z}^2-{\tt A}^2}} = i {\tt A}
$$
we obtain:
$$
\int\limits_{{\tt Z}=0}^{{\tt Z}={\tt A}} \tilde\omega_1
=\frac{\varkappa}{\sqrt{{\tt C}}}\times
\left( \frac{\pi i}{2} -\frac{i{\tt A}}{2{\tt C}} +O({\tt A}^2) \right).
$$
Therefore
$$
\varkappa = \frac{\sqrt{{\tt C}}}
{i \left( \pi -\frac{{\tt A}}{{\tt C}} \right)}+O({\tt A}^2)
$$
In the genus 1 approximation we have
$$
d\tilde p = \frac{\tt B}{2}
\frac{(E-A_0)(E-A_1)}{(E-{\tt R})\sqrt{E(E-E_1)(E-E_2)}}dE.
$$
Let us look at the conditions on $\tilde A_0$,$\tilde A_1$
imposed by
$$
\oint\limits_{\alpha_n}d\tilde p=0
$$
Expanding near ${\tt Z}=0$ we obtain
$$
d\tilde p =\frac{\tt B}{2\sqrt{\tt C}}\left(1-\frac{\tt Z}{2\tt C}
+O({\tt Z}^2)\right)
\left(\frac{({\tt Z}-\tilde A_0)({\tt Z}-\tilde A_1)}{\tt Z}\right)
\frac{d{\tt Z}}{\sqrt{{\tt Z}^2-{\tt A}^2}}=
$$
$$
=\frac{\tt B}{2\sqrt{\tt C}}\left(\frac{\tilde A_0\tilde A_1}{\tt Z} -
\left(\tilde A_0+\tilde A_1+\frac{\tilde A_0\tilde A_1}{2{\tt C}}
\right) + O({\tt Z})
\right) \frac{d{\tt Z}}{\sqrt{{\tt Z}^2-{\tt A}^2}}
$$
Due to parity constraints
$$
\oint\limits_{\alpha_n} \frac{1}{\tt Z} \frac{d{\tt Z}}{\sqrt{{\tt
      Z}^2-{\tt A}^2}}=
\oint\limits_{\alpha_n} {\tt Z} \frac{d{\tt Z}}{\sqrt{{\tt Z}^2-{\tt A}^2}}=0
$$
and finally we obtain:
\begin{equation}
\label{eq:alhas}
\tilde A_0+\tilde A_1+\frac{\tilde A_0\tilde A_1}{2{\tt C}}=O({\tt A}^2)
\end{equation}
It is difficult to use the condition
$$
\oint\limits_{\beta_n}d\tilde p=0
$$
directly. It is more convenient to use Riemann bilinear relations
instead. We know, that if
$$
\oint\limits_{\beta_n}d\tilde p= \oint\limits_{\alpha_n}d\tilde p=0,
$$
then
$$
\res[R^{\rm in}]d\tilde p = \res[\infty] \tilde p \tilde\omega_1
$$
Due to the periodicity condition
$$
\res[R^{\rm in}]d\tilde p = -i,
$$
i.e.
$$
\res[\infty] \tilde p \ \tilde\omega_1 = -i
$$
But
$$
\tilde p \tilde\omega_1 = \frac{dE}{E} \left(1+O\left(\frac{1}{E}
  \right)\right){\tt B}\varkappa
$$
therefore
\begin{equation}
\label{eq:bk}
{\tt B}\varkappa =\frac{i}{2}
\end{equation}
$$
{\tt B}=\frac{(-{\tt C})\sqrt{{\tt C}^2-{\tt A}^2}}
{{\tt C}^2+{\tt C}(\tilde A_0+\tilde A_1)+ \tilde A_0\tilde A_1}=
\frac{-{\tt C}^2}
{{\tt C}^2 +\frac{\tilde A_0\tilde A_1}{2}}+O({\tt A}^2)=
-\frac{1}
{1 +\frac{\tilde A_0\tilde A_1}{2{\tt C}^2}}+O({\tt A}^2)
$$

Direct calculation of the residues of $d\tilde p$ at the points
$R^{\rm in}$,$R^{\rm out}$,  gives us
$$
1={\tt B}\frac{\tilde A_0\tilde A_1}{ 2\sqrt{\tt C} {\tt A}}
$$
Therefore
\begin{equation}
\label{eq:def_alpha1}
\tilde A_0\tilde A_1\sim -2{\tt A}\sqrt{\tt C}
\left(1 +\frac{\tilde A_0\tilde A_1}{2{\tt C}^2} \right)
\sim -2{\tt A}\sqrt{\tt C}
\left(1 - \frac{\tt A}{{\tt C}\sqrt{\tt C}} \right)
\end{equation}
\begin{equation}
\label{eq:def_alpha2}
\tilde A_0+\tilde A_1
\sim \frac{\tt A}{\sqrt{\tt C}}
\left(1 - \frac{\tt A}{{\tt C}\sqrt{\tt C}} \right)
\end{equation}
To determine ${\tt C}$ let us use (\ref{eq:bk}).
$$
\frac{\sqrt{\tt C}} {\left( \pi -\frac{\tt A}{\tt C} \right)}\times
\frac{1}
{1 +\frac{\tilde A_0\tilde A_1}{2{\tt C}^2}} =\frac{1}{2}
$$
$$
\frac{\sqrt{\tt C}} {\left( \pi -\frac{\tt A}{\tt C} \right)}\times
\frac{1}
{1 - \frac{\tt A}{{\tt C}\sqrt{\tt C}}} =\frac{1}{2}
$$
Finally, we obtain
\begin{equation}
\label{eq:def_c}
{\tt C}=\left[\left(\frac{\pi-\frac{\tt A}{\tt C} }{2}\right)
\left(1 - \frac{\tt A}{{\tt C}\sqrt{\tt C}} \right)\right]^2
\end{equation}

To find the starting point we first solve (\ref{eq:def_c}), then
we calculate $\tilde A_0$,  $\tilde A_1$ using
(\ref{eq:def_alpha1}), (\ref{eq:def_alpha2})

Let us calculate the resonant points, defined by the condition
$$
\tilde p ({\tt C}_k) = \pi k,  \ \ k > 1.
$$
To do it let us expand $d\tilde p$ outside the neighborhood of the
point ${\tt C}$.
$$
d\tilde p =\frac{{\tt B}}{2}\frac{d{\tt Z}}{\sqrt{{\tt Z}+{\tt C}}\sqrt{{\tt Z}^2-{\tt A}^2}}
\frac{({\tt Z}-\tilde A_0)({\tt Z}-\tilde A_1)}{{\tt Z}}=
$$
$$
=\frac{{\tt B}}{2}\frac{d{\tt Z}}{\sqrt{{\tt Z}+{\tt C}}} \frac{1} {\sqrt{{\tt Z}^2-{\tt A}^2}}
\left(\frac{\tilde A_0\tilde A_1}{{\tt Z}}
  -(\tilde A_0+\tilde A_1) + {\tt Z} \right)=
$$
$$
=\frac{{\tt B}}{2}\frac{d{\tt Z}}{\sqrt{{\tt Z}+{\tt C}}} \left( \frac{1}{{\tt Z}} +O({\tt A}^2)
\right)
\left(\frac{\tilde A_0\tilde A_1}{{\tt Z}}
  -(\tilde A_0+\tilde A_1) + {\tt Z} \right)=
$$
$$
=\frac{{\tt B}}{2}\frac{d{\tt Z}}{\sqrt{{\tt Z}+{\tt C}}}
\left( 1 -\frac{\tilde A_0+\tilde A_1}{{\tt Z}} +
\frac{\tilde A_0\tilde A_1}{{\tt Z}^2} + O({\tt A}^2) \right)=
$$
$$
=\frac{{\tt B}}{2}\frac{dE}{\sqrt{E}}\left( 1 -
\frac{\tilde A_0+\tilde A_1}{E-{\tt C}} +
\frac{\tilde A_0\tilde A_1}{(E-{\tt C})^2} + O({\tt A}^2) \right)
$$
Using the table integrals:
$$
\int \frac{dE}{2\sqrt{E}} = \sqrt{E}
$$
$$
\int \frac{dE}{2\sqrt{E}(E-{\tt C})} = \frac{1}{2\sqrt{{\tt C}}}
\log\left(\frac{\sqrt{E}-\sqrt{{\tt C}}}{\sqrt{E}+\sqrt{{\tt C}}}\right)
$$
$$
\int \frac{dE}{2\sqrt{E}(E-{\tt C})^2} = -\frac{1}{4{\tt C}}
\left[\frac{2\sqrt{E}}{E-{\tt C}}
+\frac{1}{\sqrt{{\tt C}}}
\log\left(\frac{\sqrt{E}-\sqrt{{\tt C}}}{\sqrt{E}+\sqrt{{\tt C}}}\right)\right]
$$
we get
$$
\tilde p = {\tt B}\left[ \sqrt{E} -\left(\tilde A_0+\tilde A_1+
\frac{\tilde A_0\tilde A_1}{2{\tt C}}\right)  \frac{1}{2\sqrt{{\tt C}}}
\log\left(\frac{\sqrt{E}-\sqrt{{\tt C}}}{\sqrt{E}+\sqrt{{\tt C}}}\right)-
\frac{\tilde A_0 \tilde A_1}{2{\tt C}}\frac{\sqrt{E}}{E-{\tt C}}+O({\tt A}^2)
\right]
$$
Taking into account (\ref{eq:alhas}) we obtain
$$
\tilde p = {\tt B}\sqrt{E} \left[ 1 -
\frac{\tilde A_0 \tilde A_1}{2{\tt C}}\frac{1}{E-{\tt C}}+O({\tt A}^2)
\right]
$$
Therefore we calculate the resonant points from the equation
\begin{equation}
{\tt C}_k = \left[\frac{\pi k}{{\tt B}}\frac{1}{ 1 -
\frac{\tilde A_0 \tilde A_1}{2{\tt C}({\tt C}_k-{\tt C})}}\right]^2
\end{equation}

\end{document}